\documentclass[aps,prb,showpacs,twocolumn]{revtex4}
\usepackage{graphicx}

\newcommand{\be}{\begin{eqnarray}}
\newcommand{\ee}{\end{eqnarray}}
\newcommand{\bmat}{\left(\begin{array}}
\newcommand{\emat}{\end{array}\right)}
\newcommand{\no}{\nonumber}
\newcommand{\e}{\epsilon}

\begin{document}
\title{Mesoscopic conductance fluctuations in a coupled quantum dot system}
\author{Kazutaka Takahashi}
\affiliation{Department of Physics, Tokyo Institute of Technology,  
 Tokyo 152--8551, Japan}
\affiliation{Condensed Matter Theory Laboratory, RIKEN, 
 Saitama 351--0198, Japan}
\author{Tomosuke Aono}
\affiliation{Department of Physics, Ben--Gurion University of the Negev,
 Beer--Sheva 84105, Israel}
\date{\today}

\begin{abstract}
 We study the transport properties of 
 an Aharonov--Bohm ring containing two quantum dots.
 One of the dots has well--separated resonant levels, while
 the other is chaotic and is treated by random matrix theory.
 We find that the conductance through the ring 
 is significantly affected by mesoscopic fluctuations.
 The Breit--Wigner resonant peak is changed to 
 an antiresonance by increasing the ratio of the level broadening 
 to the mean level spacing of the random dot.
 The asymmetric Fano form turns into a symmetric one 
 and the resonant peak can be controlled by magnetic flux.
 The conductance distribution function clearly shows 
 the influence of strong fluctuations.
\end{abstract}
\maketitle


 The quantum dot~\cite{Beenakker, Alhassid} (QD) is an ideal system to 
 study the phase coherence of quantum mechanical wave functions.
 Such effects can be explained by the interference of different pathways, 
 induced by elastic scattering from an irregular boundary and/or
 impurities. 
 When a dot is connected to leads, there is a strong overlap between 
 the dot and leads and we must treat the whole as a quantum system.
 In fact, a recent numerical calculation of a chaotic dot 
 shows that there are scars that connect between the leads.~\cite{SI}

 The coexistence of a direct path and discrete levels 
 in the dot induces a prominent effect, the Fano effect.~\cite{Fano}
 It was shown in QD systems~\cite{KAKI, JMHG} that asymmetric Fano peaks 
 can be controlled by gate voltages and magnetic fields.
 The work in Ref.~\onlinecite{KAKI} demonstrated that  an Aharonov-Bohm 
 (AB) ring is a suitable system to study this effect.
 A dot is embedded in one of the arms, 
 and the ring geometry is utilized as an interferometer.
 This effect has also been observed in other experiments, including
 a microwave cavity~\cite{RLBKS} and optical absorption.~\cite{FCSWP}

 The situation changes drastically 
 if we take into account mesoscopic fluctuations.
 A random distribution of levels in the dot leads to 
 sample--to--sample fluctuations of the conductance.
 In contrast to bulk systems, such fluctuations cannot be neglected 
 in QD systems, and the conductance distribution has 
 a broad non--Gaussian shape.~\cite{PEI}
 It was observed in a QD system~\cite{HPMBDH} 
 and in a chaotic microwave cavity.~\cite{HZOAA}
 Clerk {\it et al.} developed a statistical theory 
 of Fano peaks assuming a random distribution of peaks.~\cite{CWB}
 These authors focused on resonance--to--resonance fluctuations of 
 the asymmetric Fano form, rather than conductance fluctuations.

 In this paper we develop a statistical theory 
 of the AB ring with two QDs in the arms.
 In addition to a resonant dot in one of the arms, 
 a random dot is connected to the other arm 
 and is treated by random matrix theory (RMT).~\cite{Mehta} 
 The orbit through the resonance is correlated to 
 those through the random levels, 
 and strong fluctuations can be observed in the conductance.
 We show that the Breit--Wigner and Fano resonant forms 
 are no longer maintained in the averaged conductance.

 We focus on the following two issues.
 First, we examine the properties of those orbits that 
 contribute to the conductance.
 Motivated by the work of Ref.~\onlinecite{CWB}, we take into account 
 the direct nonresonant path in the random dot as well as 
 the Breit--Wigner resonant path in the regular dot.
 We show that these two channels give qualitatively different 
 contributions to the conductance in the presence of random levels.
 Second, the level broadening of the random dot.
 For an open dot, the broadening can be larger than 
 the mean level spacing due to strong coupling to the 
 leads.~\cite{Beenakker, Alhassid}
 We systematically change the broadening from small to large values
 to elucidate how this parameter affects the results.

\begin{figure}[tb]
\begin{center}
 \includegraphics[width=0.9\columnwidth]{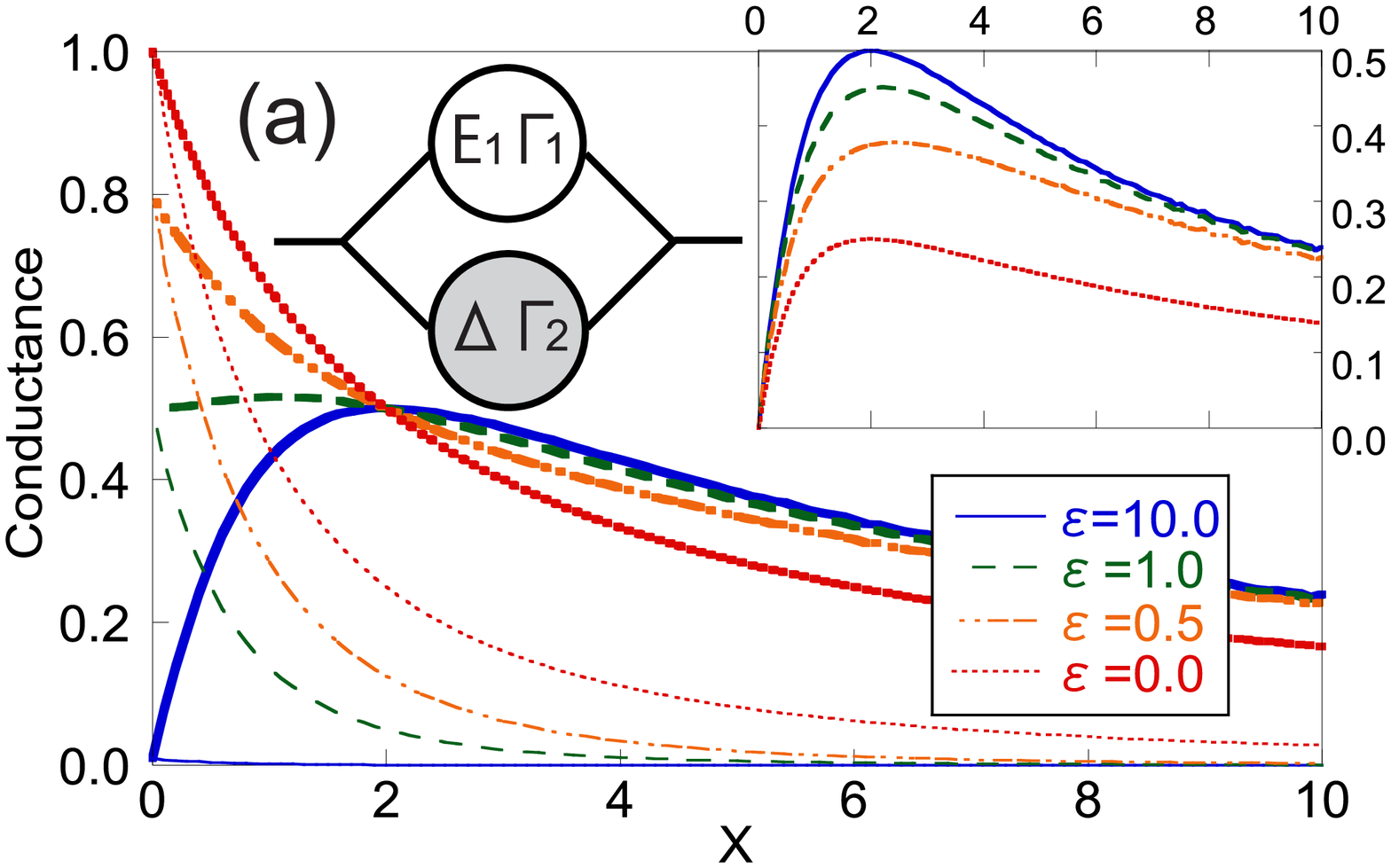} 

 \medskip

 \includegraphics[width=0.9\columnwidth]{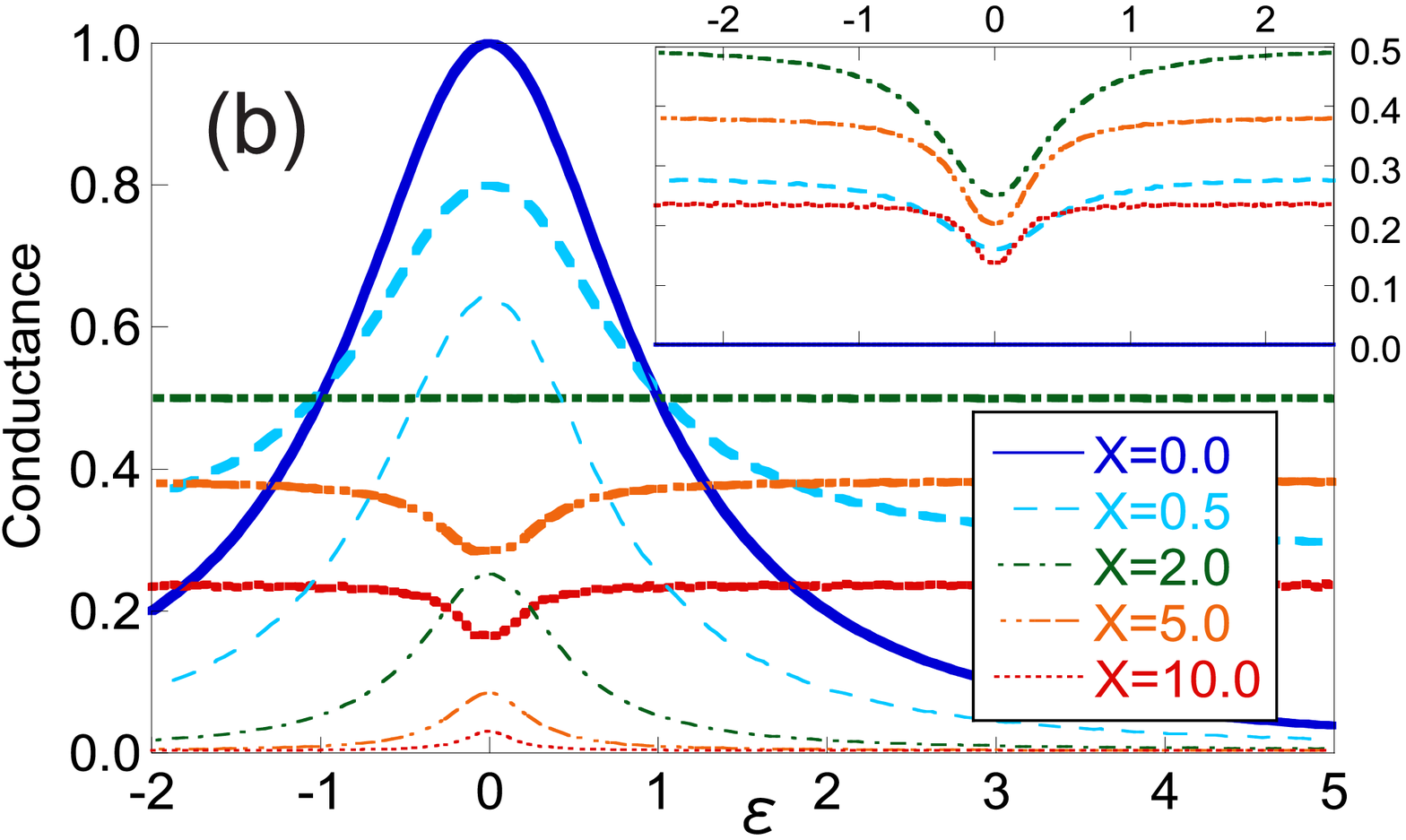}
 \caption{(Color Online)
 Conductance vs $X=\pi\Gamma_2/\Delta$ (a) 
 and $\e=(E-E_1)/\Gamma_1$ (b) at $a=0$. 
 The thick (thin) lines are the total conductance $g$ 
 (the principal part $g_0$).
 Upper left inset in (a): Sketch of the sample. 
 The parameters characterizing each dot are shown in the figure.
 Upper right inset in (a) and (b): Fluctuation part $\delta g=g-g_0$.
}
 \label{xeg}
\end{center}
\end{figure}

 Our system is defined by the internal Hamiltonian 
 for the two QDs and their couplings to the left and right leads.
 It is depicted in the upper left inset in Fig.~\ref{xeg}(a).
 We assume that one of the dots (dot 1) has regular resonant levels 
 and the level spacing is much larger than the level broadening.
 In this case, each of the levels can be treated independently.
 The other dot (dot 2) is relatively large and 
 has many levels distributed randomly.
 As is known from scattering theory 
 (see, e.g., Refs.~\onlinecite{Beenakker, Alhassid}, 
 and \onlinecite{VWZ}),  
 the $S$ matrix of the system is written as 
\be
 S = 1-2\pi i w^\dag\frac{1}{E-H+i\pi ww^\dag}w, \label{S}
\ee
 where $H=H_1\oplus H_2$ denotes the Hamiltonian 
 for the QDs, $w$ is the dot--lead coupling, and 
 $E$ is the (Fermi) energy.
 We adopt a single--level Hamiltonian $H_1=E_1$ for dot 1.
 $H_2$ for dot 2 is a member of the Gaussian unitary 
 ensemble~\cite{Mehta} and its size $N$ 
 is taken to be infinity to find the universal limit.
 We assume that the $S$ matrix is a $2\times 2$ matrix, 
 which means the left and right leads have a single channel respectively.
 The conductance, measured in units of $2e^2/h$, 
 is calculated from $g = \langle |S_{12}|^2\rangle$.

 In the present RMT approach, 
 there is no need to know the full form of $w$.
 The matrix $w$ appears in the $S$ matrix formula (\ref{S})
 as $w^\dag G w$, where $G=(E^+-H)^{-1}$.
 After the averaging, the matrix structure of $G$ is lost and 
 $2\times 2$ matrices $\gamma_{i} = \pi w^{(i)\dag} w^{(i)}$ 
 appear in averaged quantities.
 Here $i=1,2$ label the dot and $w^{(1)}$($w^{(2)}$) is 
 a $1\times 2$ ($N\times 2$) matrix.
 Assuming a symmetric coupling with respect to the left and right leads, 
 we arrive at the general form
\be
 \gamma_1 = \frac{\Gamma_1}{2}\bmat{cc} 1 & e^{i\varphi} \\
 e^{-i\varphi} & 1 \emat, \ 
 \gamma_2 = \frac{N\Gamma_2}{2}\bmat{cc} 1 & ae^{-i\varphi} \\
 ae^{i\varphi} & 1 \emat, 
\ee
 where $\Gamma_i$ is the level broadening of the dot $i$
 and $\varphi$ is the AB flux.~\cite{GIA-KK}
 The real parameter $a$ ($0\leq a\leq 1$) represents the strength of 
 the nonresonant direct path since the off--diagonal part of $\gamma$ 
 contributes to the conductance directly.~\cite{VWZ}


 The averaged S matrix is calculated from the relation
 $(1-\langle S\rangle)/(1+\langle S\rangle)
 =(1-S_1)/(1+S_1)+(1-\langle S_2\rangle)(1+\langle S_2\rangle)$, 
 where $S_{1,2}$ are the $S$ matrices of the dot $1,2$ respectively.
 In RMT the magnitude of the Gaussian fluctuations 
 determines the mean level spacing $\Delta$.
 Then the averaged Green's function of the dot 2 is given by  
 $\langle G\rangle = -i\pi/N\Delta$, 
 where we take the limit $N\to\infty$ and neglect the real part of $G$.
 In this limit we obtain
 $\langle S_2\rangle = (1-\pi\gamma_2/N\Delta)/(1+\pi\gamma_2/N\Delta)$
 while $S_1=(E-E_1-i\gamma_1)/(E-E_1+i\gamma_1)$.
 $\langle S \rangle$ is determined by the following four parameters:
 $a$, $\varphi$, $\e = (E-E_1)/\Gamma_1$, and $X = \pi\Gamma_2/\Delta$.
 $\e$ measures the distance from the resonance point of the regular dot
 and $X$ represents the ratio of the level broadening to 
 the level spacing for the random dot.
 When $X \gg 1$, the level spectrum is continuous.
 The use of RMT implies our consideration is restricted 
 to the energy scale much smaller than the Thouless energy.

 If we disregard quantum fluctuations, the conductance is given by
 $g_0=|\langle S_{12}\rangle|^2$, 
 which we call the principal part of $g$.
 It is given by the Fano form 
\be
 g_0 &=& \frac{a^2 X^2}{\left[1+(1+a)X/2\right]^2
 \left[1+(1-a)X/2\right]^2} \no\\
 & & \times\frac{\left|E-E_1+q\Gamma_1\right|^2}
 {(E-E_1)^2
 +\frac{\left[1+(1-a \cos 2\varphi)X/2\right]^2}
 {\left[1+(1+a)X/2\right]^2
 \left[1+(1-a)X/2\right]^2}\Gamma_1^2}, \label{g0}
\ee
 where $q = ie^{-2i\varphi}/a X$ is the Fano parameter.
 This parameter is complex and becomes purely imaginary 
 when the AB flux $\varphi$ is zero, 
 which is contrasted with the experiments for clean 
 systems~\cite{KAKI, JMHG} where a real $q$ has been 
 observed in the absence of a magnetic field.
 We also note that the result of Eq.(\ref{g0}) holds 
 regardless of the choice of the universality class 
 because the resonance is not treated 
 randomly, in contrast with the approach in Ref.~\onlinecite{CWB}.
 When $a=0$, $|q|\to\infty$ and the result reduces to
 the standard Breit--Wigner form.
 The presence of the random dot leads to a reduction of the level 
 broadening and the conductance by a factor of $1/(1+X/2)^2$.
 

 We now consider mesoscopic fluctuations of the conductance 
 $\delta g = g-g_0$.
 We calculate these using the method of supersymmetry,~\cite{VWZ} 
 which allows us to derive the nonlinear $\sigma$ model
\be
 F = \frac{1}{2}\mbox{str}_8 \ln\left(1+\frac{T/2}{1-T/2}
 \frac{\Lambda\sigma+\sigma\Lambda}{2}\right), 
\ee
 where the $4\times 4$ supermatrix $\sigma$ parametrizes 
 the saddle point manifold and satisfies $\sigma^2=1$.~\cite{Efetov}
 $\mbox{str}$ denotes the supertrace 
 and $\Lambda=\mbox{diag}(1,-1)$ in retarded--advanced space.
 The $2\times 2$ matrix $T$ defined by 
 $T = 1-\langle S\rangle\langle S\rangle^\dag$
 is called the transmission coefficients.
 This form of the $\sigma$ model is well known
 as a standard model of a single random dot.
 The only difference is in the $T$ matrix.
 The fact that the $\sigma$ model is written 
 in terms of $T$ only demonstrates the universality of 
 the correlation functions of the S matrix elements.
 For instance, $\delta g$ in Eq.(\ref{deltag}) is a function of $T$.
 This is to be contrasted with the result of Eq.(\ref{g0}), 
 where such $T$ universality does not hold.
 In the present model, we find the eigenvalues of $T$ at $a=0$ 
\be
 T_1 = \frac{2X}{\left(1+X/2\right)^2}, \qquad
 T_2 = \frac{2X}{\left(1+X/2\right)^2+1/\e^2}. \label{T12}
\ee
 We see that $T_1$ is independent of $\e$.
 

 We present the analytical results for the conductance
 together with numerical ones. 
 Numerical calculations are performed 
 using the random $S$ matrix model,~\cite{randomS}
 where the internal structure of the $S$ matrix is disregarded and 
 randomness is imposed directly on $S$. 
 For the distribution of $S$, we use the generalized circular unitary 
 ensemble (CUE) based on the Poisson kernel~\cite{pker} 
\be
 P(S)d\mu = 
 \frac{\det(1-\langle S\rangle\langle S\rangle^\dag)^2}
 {\left|\det(1-S\langle S\rangle^\dag)\right|^4} d\mu,
\ee
 where $d\mu$ is the measure of the CUE.~\cite{Mehta}
 This is the maximally randomized distribution 
 under the condition that the average value is $\langle S\rangle$.
 Following the previous approach for regular systems~\cite{GIA-KK}
 we separate the system into the dots 1 and 2, 
 and the left and right forks that connect the dots to leads.
 Choosing the fork $S$ matrix in a symmetric form 
 we find the transmission through the ring expressed 
 by the $S$ matrix $S_{1,2}$ for each dot.
 $S_2$ is treated statistically using a Poisson kernel.
 This random $S$ matrix approach is equivalent to  
 the random Hamiltonian approach if $\langle S\rangle$ is
 chosen properly and $N\to\infty$.~\cite{Brouwer}
 We use the same expressions for $S_1$ and $\langle S_2\rangle$ 
 as in the random Hamiltonian approach.

 First we consider the $X$ dependence of the conductance at $a=0$.
 When $\e\to\infty$, where the regular dot is detached 
 from the random dot, we have $T_1=T_2=2X/(1+X/2)^2$, 
 and recover the known result~\cite{Efetov2}
\be
 g_0=0, \quad \delta g = \frac{T_1}{3}+\frac{T_1^2}{6}.
 \label{deltag}
\ee
 At the resonance point $\e=0$,
 $T_1=2X/(1+X/2)^2$, $T_2=0$, and we obtain 
\be
 g_0 = \frac{1}{\left(1+X/2\right)^2}, \quad 
 \delta g = \frac{T_1}{4}=\frac{X/2}{\left(1+X/2\right)^2}. \label{grp}
\ee
 In Fig.~\ref{xeg}(a), $X$ dependence of the conductance 
 is shown for several values of $\e$.
 $g_0$ shows a peak at $X=0$
 while $\delta g$ takes a maximum at $X=2$, as shown by 
 the thin lines and the inset in Fig.~\ref{xeg}(a), respectively.
 As $\e\to\infty$ 
 $g_0$ ($\delta g$) is monotonically decreasing (increasing) 
 and the result rapidly approaches Eq.(\ref{deltag}).

 $\e$ dependence of the conductance is shown in Fig.~\ref{xeg}(b).
 A resonance peak appears at $\e=0$, 
 reflecting transport through the regular dot.
 This peak structure, however, changes qualitatively as a function of $X$.
 For small $X$ the peak is convex and 
 the peak height decreases on increasing $X$.
 When $X=2$, $g$ is independent of $\e$.
 Increasing $X$ further, we find that the peak turns into 
 an antiresonance and $g$ decreases monotonically.
 The result for $X=2$ corresponds to that of 
 the CUE because $\langle S_2\rangle=0$, and 
 the Poisson kernel $P(S_2)$ becomes unity.

\begin{figure}[tb]
\begin{center}
 \includegraphics[width=0.9\columnwidth]{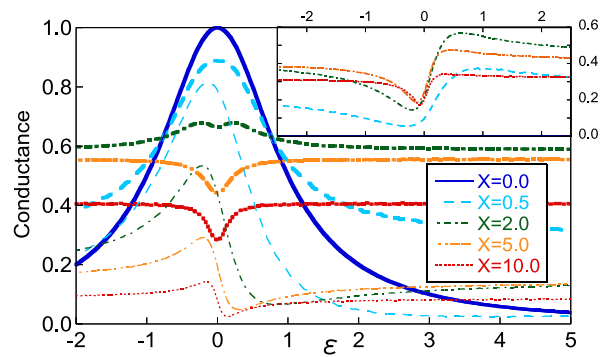}
 \caption{(Color Online)
 Conductance vs $\e$ at $a=0.7$ and $\varphi=\pi/8$.
 The thick (thin) lines are the total conductance $g$
 (principal part $g_0$).
 Inset: Fluctuation part $\delta g$.}
 \label{fanog}
\end{center}
\end{figure}


 The obtained results are for a model with a resonance at $a=0$. 
 Now the question is whether they are characteristic of
 the resonance model.
 For comparison, we discuss the direct reaction model 
 with $a\ne 0$ and $\e\to\infty$.
 Both models are similar in that 
 the conductance $g_0$ becomes finite in the absence of fluctuations.
 $g_0$ becomes maximum at $a=1$, 
 which is similar to the situation at $\e=0$ in the resonance model.
 In the direct reaction model, 
 the eigenvalues of the $T$ matrix are given by 
\be
 T_{1,2} = \frac{2(1\pm a)X}{\left[1+(1\pm a)X/2\right]^2}.
\ee
 $T_2$ goes to zero as $a\to 1$ as in the limit $\e\to 0$ in Eq.(\ref{T12}). 
 In this limit, we find $T_1=4X/(1+X)^2$, $T_2=0$, and 
\be
 g_0 = \frac{X^2}{\left(1+X\right)^2}, \quad 
 \delta g = \frac{T_1}{4}=\frac{X}{\left(1+X\right)^2}.
\ee
 Note that $T_1$ dependence of $\delta g$ is the same as in Eq.(\ref{grp}) 
 but the $X$ dependence is different.
 We have found numerically that 
 the peak is maintained for an arbitrary value of $X$ and 
 that there is no antiresonance, in contrast to the resonance model.


\begin{figure}[b]
\begin{center}
 \includegraphics[width=0.9\columnwidth]{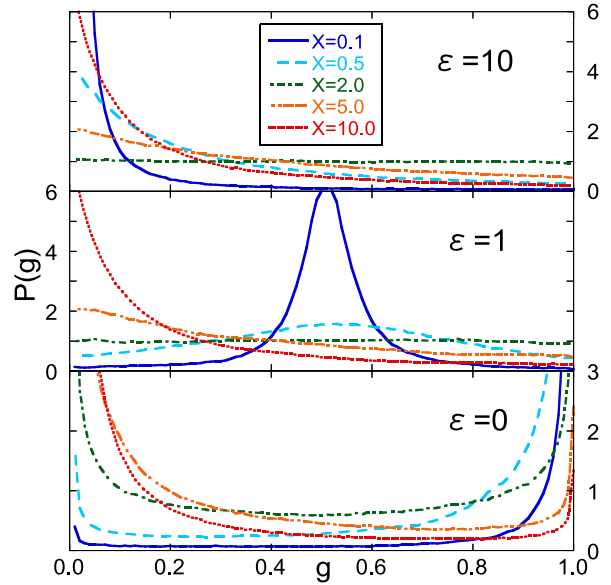}
 \caption{(Color Online)
 Conductance distribution at $a=0$.}
 \label{pg}
\end{center}
\end{figure}

 As we have shown in Eq.(\ref{g0}), when we consider both the resonance 
 and direct reaction channels, the Fano effect appears in $g_0$.
 In Fig.~\ref{fanog}, a typical numerical result of the conductance is shown 
 at $a=0.7$ and $\varphi=\pi/8$.
 We observe that an asymmetric form is obtained for the fluctuation part 
 as well as for the principal part.
 However, these asymmetric peaks work to compensate each other,
 resulting in a symmetric $g$ 
 that has a form similar to the previous result at $a=0$.
 The difference is that the conductance as a whole is enhanced 
 due to the direct path contribution.
 Far from the resonance point, 
 the conductance is independent of $\varphi$,
 while it is sensitive at the resonance point.
 The resonance (antiresonance) is amplified at $\varphi=0$ ($\pi/4$)
 where the Fano parameter $q$ is pure imaginary (real).
 When $q$ takes a complex value (when $0<\varphi<\pi/4$), a dip 
 in the resonance peak is formed at the intermediate values of $X$ 
 (See the plot of $X=2$ in Fig.~\ref{fanog}).


 As we have shown, $\delta g$ are of the order of $g_0$ and
 the mere calculation of the averaged conductance is not enough 
 to characterize the system.
 We show the numerical results for the conductance distribution function  
 $P(g)=\langle\delta (g-|S_{12}|^2)\rangle$ in Fig.~\ref{pg}.
 For $\e\gg 1$ we find the results in Refs.~\onlinecite{PEI} and 
 \onlinecite{randomS} expressed in terms of $T_1$ in Eq.(\ref{T12}).
 For finite value of $\e$, the $T$ universality does not hold and 
 the results essentially depend on $X$ and $\e$.
 When $X<2$, the peak representing the resonant conductance 
 moves from $g=0$ to 1 as $\e$ decreases.
 When $X>2$, the resonant conductance is suppressed 
 and the distribution function shows 
 the strong influence of the chaotic scattering.
 When $\e=0$, two peaks appear at $g=0$ and $1$.
 The peak at $g=0$ is larger (smaller) than that at $g=1$ 
 when $X>2$ ($X<2$), which clearly shows the coexistence of 
 the contributions from both regular and chaotic dots.
 At $X=2$, $P(g)$ is always symmetric and
 is consistent with the results of the averaged conductance.
 The curve for $\e=0$ is well fitted by the function 
 $P(g)=1/\pi\sqrt{g(1-g)}$.


 We remark on the effect of dephasing.
 It can be simply studied by introducing 
 an imaginary part to the energy $E\to E+i/2\tau$.
 The substitution of this to Eq.(\ref{g0})
 leads to the reduction of the conductance.
 The effect on the fluctuation part is to add 
 $F_\tau = (\Delta\tau)^{-1}\mbox{str}\,\sigma\Lambda$
 to the $\sigma$ model.
 Numerically we have observed that the fluctuation part is strongly 
 suppressed by this effect while the principal part shows 
 a small reduction.
 This means that the resonance is preserved at any $X$.
 We also anticipate that an asymmetric Fano resonance can be observed, 
 since the cancellation of the asymmetry between 
 the principal and fluctuation parts becomes incomplete.
 The quantitative estimate of the dephasing effects 
 using the method in Ref.~\onlinecite{BB}
 is necessary to compare with the experimental results, as was 
 done in Refs.~\onlinecite{HPMBDH} and \onlinecite{HZOAA} 
 for the single--dot systems.
 A detailed study will be reported elsewhere.


 Another interesting problem 
 is the ensemble dependence of the result.
 Our numerical calculations using the orthogonal and symplectic 
 ensembles show that the distribution at 
 the resonant point (the lowest graph in Fig.~\ref{pg}) is 
 independent of the ensemble.
 It implies the distribution is determined by some universal mechanism.


 In conclusion, we have developed a statistical theory 
 for an AB ring system with regular and chaotic QDs.
 The conductance and its distribution are strongly influenced by 
 the mesoscopic fluctuations of the chaotic dot and 
 the position of the resonance peaks.

 We are grateful to A. Furusaki and S. Iida for useful discussions.
 T.A. acknowledges support from 
 the Japan Society for the Promotion of Science.


\end{document}